# News Sentiment as Leading Indicators for Recessions

Melody Y. Huang, Randall R. Rojas, Patrick D. Convery

Department of Economics

University of California, Los Angeles

Los Angeles, CA 90095

May 30, 2018

## Abstract

In the following paper, we use a topic modeling algorithm and sentiment scoring methods to construct a novel metric that serves as a leading indicator in recession prediction models. We hypothesize that the inclusion of such a sentiment indicator, derived purely from unstructured news data, will improve our capabilities to forecast future recessions because it provides a direct measure of the polarity of the information consumers and producers are exposed to. We go on to show that the inclusion of our proposed news sentiment indicator, with traditional sentiment data, such as the Michigan Index of Consumer Sentiment and the Purchasing Manager's Index, and common factors derived from a large panel of economic and financial indicators helps improve model performance significantly.

# 1 Introduction

Traditional recession prediction literature has relied heavily on finding economic and financial indicators that hold predictive power in forecasting or explaining recessionary periods. For example, numerous papers have studied the predictive capabilities of the yield curve spread between the ten-year treasury rate and the three-month treasury rate (Estrella, 2005, Estrella and Trubin, 2006, Calvao, 2007). Rudebusch and Williams, 2009 demonstrated that in the context of a real-time forecasting model, the yield spread actually outperforms professional forecasters in forecasting future recessions. Other predictive factors that have been studied largely center on macroeconomic and financial indicators, such as latent factors, like the slope, level, and curvature of the yield curve (Ang and Piazzesi, 2003, Diebold, et al. 2006, Bordo and Haubruch, 2008), spreads of different treasury rates (Ang, et al., 2006), autoregressive properties of GDP (Ang, et al., 2006), unemployment and stock price indices (Chionis, et al., 2009), to name a few.

Moench, 2008 proposed using the common components of macroeconomic variables in order to estimate a model, allowing for vast improvements in the short-term forecast errors. In picking the optimal combination of variables, Liu and Moench, 2016 and Ng, 2014 evaluated large panels of traditional macroeconomic and financial indicators in their predictive capabilities. However, in general, the conclusion is that while the introduction of a large panel of macroeconomic and financial variables can be helpful to improving forecasting accuracy, the majority of the predictive power from the proposed recession models derives from the yield curve, or a combination of the yield curve, the short-term rate, and the stock market (Estrella and Mishkin, 1998, Wright, 2006, Nyberg, 2010, Lahiri, et. al 2013, Croushore and Masten, 2014, Liu and Moench, 2016).

The importance of understanding sentiment in the context of macroeconomic activity in order to successfully forecast downturns in business cycles has been widely discussed in literature. Sentiment indicators are crucial to understanding recessions because they capture a sense of animal spirits (Keynes, 1936), while others have argued that sentiment indicators are a proxy for how people and other agents within an economic system understand and process news of changes within the economy as a whole (Christiansen, et. al 2014, Barksy and Sims, 2012). From a very fundamental level, a sentiment indicator can account for the fact that simply the perception of how the economy is doing can influence the actual performance of the economy (Bernanke 2008). Additionally, sentiment indicators have the advantage over traditional leading indicators, in that while macroeconomic variables are often published with some delay and require future revisions, sentiment indices are available in real-time (Koenig, 2002).

Matsusaka and Sbordone, 1995 and Batchelor and Dua, 1998 demonstrated that the inclusion of consumer sentiment could drastically help improve forecasts of GDP. Christiansen, et. al 2014 demonstrated that sentiment variables were found to significantly improve both in-sample and out-of-sample performance of recession prediction models, especially when paired with common recession predictors and factors. Additionally, the combination of such indicators with classical recession predictors mitigates prior concerns of structural breaks and instabilities present in previous recession prediction models (Stock and Watson, 2003; Ng and Wright, 2013; Rossi, 2013; Christiansen, 2014). Barksy and Sims, 2012 argue that it is not necessarily confidence that is causal in inducing changes in economic activity, but rather the fundamental news that induces changes in consumer and business confidence. Christiansen, et. al 2014 made the argument that business confidence reacted faster to fundamental news, and therefore serves as a stronger indicator.

Previous attempts at studying sentiment have relied upon survey data. Many studies (such as Christiansen, et. al 2014, Howry, 2001, and Barksy and Sims, 2012 to name a few) have used University of Michigan's Index of Consumer Sentiment and the Institute of Supply Management's Purchasing Managers Index as a measure for sentiment. Our paper seeks to add to the existing recession prediction literature by studying the potential for using not only consumer and business confidence as a measure for sentiment, but also the underlying sentiment of news itself.

With the advent of big data, there has been an increasing usage of unstructured data as potential sentiment predictors in forecasting models. For example, Bollen, et al. 2011 found that applying sentiment analysis across Twitter feeds could help predict directional stock market movements, while Hisano, et al. used topic modeling across news articles to forecast market volatility. However, much of the work in using unstructured data, such as social media, internet searches, and news sentiment, as factors within finance and economic models have focused on predicting financial markets. D'Amuri and Marcucci, 2017 demonstrated that there existed potential for using Google searches in forecasting unemployment rates in the United States, and Choi and Varian, 2012 used Google Trends data to perform short-term forecasts of economic indicators, such as automobile sales, unemployment claims, and travel destination planning.

Our paper draws upon similar methods to create a novel news sentiment factor for predicting recessions. We construct a sentiment-based indicator that represents the polarity of economic news at a given point in time using a combination of traditional sentiment analysis and topic modeling algorithms. This is then included with traditional sentiment indicators derived from survey data, as well as a variety of previously identified predictive macroeconomic and financial time series indicators. We show that even in the presence of all of these factors, our novel metric is consistently statistically significant in its predictive capabilities.

Additionally, using a similar evaluative framework as Berge and Jord, 2011 and Liu and Moench, 2016 with AUROC and ROC curves, we find that the combination of the survey-based indicators with our proposed indicator show larger degrees of predictive power than a model consisting of traditional, macroeconomic variables. We hypothesize that this may be due to the fact that it is a combination of how the inherent sentiment of economic news (quantified by our indicator), as well as how the public responds (quantified by survey data), that affect the probability of the economy entering a recessive state.

The paper is structured as follows. In section 2, we explain the intuition behind how we construct our novel sentiment indicator. In sections 3, we detail our model specification and introduce the idea of interaction terms between the different sentiment indicators. In section 4, we examine both in-sample and out-of-sample performances of the different models and find that at longer time horizons, including our novel sentiment indicator can help improve forecasting capabilities significantly.

## 2 Constructing the Indicator

Our goal is to provide a measure for the overall sentiment of the news. We examine two components of news that may affect the public perception of an event. The first is the general positive and negative nature of the articles being published. With more negative news being published about the economy, we would expect that the overall public perception of the economy would be negative. The second is the overall concentration of topics in the news. When financial markets are performing poorly, or when the economy is in a recession, the majority of news stories will be dedicated to the downturn in the market. However, when the markets are performing well, or the economy is an upswing, news stories related to financial markets will be more sparse in topics.

We begin by scraping all articles published in the New York Times from the years 1965 to 2017 related to the economy and financial markets. This is done by conditioning the search for published articles linked the keywords "economy" and "stock market." Per standard text mining practice, numbers, punctuation marks, white spaces, and common stop words are removed from each article. The articles are then sorted by date of publication and subset into monthly blocks.

## 2.1 Scoring the Sentiment

The overall positivity or negativity of the news in a given time period is measured by first scoring each word in a given document using an opinion lexicon (see Liu, 2008). Words are either given a score of +1, -1, or 0 for positive, negative, or neutral sentiment. The total score of each document is then computed by summing up the overall sentiment in a given document and then dividing by the total number of words. This is done for all documents, and the scores for each document in a given day are summed together for a "total" sentiment score for finance related news in a given day. The reason a sum is used instead of an averaging is that we want to account for the fluctuations in the number of stories being published in a given day. While more complex methods could have been employed to measure sentiment in a more accurate manner (i.e., certain lexicons allow for multidimensionality in sentiment, with more possibilities to allow for variation in mood than just a simple +1/-1 scoring metric for positive/negative words), we opt for this two dimensional concept of sentiment for simplicity and interpretability.

## 2.2 Measuring the Concentration of Topics

To construct a proxy for the relatedness of articles to one another in a particular time period, we use the Latent Dirichlet Allocation (LDA) topic modeling algorithm to estimate the underlying topic distributions in a body of articles. LDA assumes a generative process in which documents are represented as a distribution over topics, and topics are distributions over words. A corpus of $M$ documents is said to contain $V$ unique words (known as 'the vocabulary' of a given corpus), and each of the $M$ documents is assumed to contain a sequence of words $\{w_{1_d}, w_{2_d}, ..., w_{N_d}\}$. Therefore, the process follows as such: given a set number of topics $K$ within a corpus of documents, $\theta_i$ is drawn from a Dirichlet distribution with parameter $\alpha$, where $i \in \{1, ..., M\}$. $\phi_k$ is also drawn from a Dirichlet distribution with parameter $\beta$, where $k \in \{1, ..., K\}$. Then, for each word position $i, j$ (where $j \in \{1, ..., N\}$, and $N_i$ represents the length of the $i$-th document), a topic $z_{i,j}$ is taken from a multinomial distribution with parameter $\theta_i$, and a word $w_{i,j}$ is taken from a multinomial distribution with parameter $\phi_{z_{i,j}}$ [11].

The LDA model specifies the following probability distribution over words within a given document:

$$P(w_i) = \sum_{j=1}^{T} P(w_i|z_i = j)P(z_i = j), \tag{1}$$

where $P(z_i = j)$ represents the probability that the $j$-th topic was sampled for the $i$-th word, and $P(w_i|z_i = j)$ is the probability of word $w_i$ under topic $j$ [11, Steyvers et al. 2007]. In reality, we are only able to observe individual words from our documents (i.e., $w_{i,j}$); all other variables are latent. We are interested in the posterior multinomial distribution of $\phi$, a $K \times V$ Markov matrix, where $V$ represents the dimension of the vocabulary [11]. $\phi^j$, therefore, is analogous to $P(w|z = j)$. While the optimal number of topics $K$ for each corpus of documents varies, varying the input of $K$ did not alter the results much. Therefore, while there exist different methods to determine the best number of topics $K$ to fit the model with (i.e., Cao, et al. 2009, as an example), we fix $K$ at a set number of 30 for simplicity.

Our goal is to find a metric to represent the relatedness of the topics estimated within our corpus of documents. A commonly used metric to measure the differences between two probability distributions is the Kullback-Leibler divergence [Sievert et al. 2014]. Given two discrete probability distributions, $P$ and $Q$, the Kullback-Leibler divergence of $Q$ from $P$ is defined as:

$$D_{KL}(P||Q) = \sum_i P(i) \log \frac{P(i)}{Q(i)} \tag{2}$$

A similar method based on the Kullback-Leibler divergence, the Jensen-Shannon divergence, is finitely bound and symmetric. It is used traditionally to compute similarities between two distributions. The Jensen-Shannon divergence of $Q$ from $P$ is therefore:

$$JSD(P||Q) = \frac{1}{2}D(P||M) + \frac{1}{2}D(Q||M), \tag{3}$$

where $M = \frac{1}{2}(P+Q)$. Taking the square root of the Jensen-Shannon divergence gives us the Jensen-Shannon distance. We use the estimated probability distributions for the topics extracted by the LDA algorithm ($\phi$) and compute the Jensen-Shannon distance between each topic. This returns a $K \times K$ distance matrix of the form:

$$M_{dist} = \begin{bmatrix} 0 & d_{1,2} & d_{1,3} & \dots & d_{1,k} \\ d_{2,1} & 0 & d_{2,3} & \dots & d_{2,k} \\ \vdots & \vdots & \vdots & \ddots & \vdots \\ d_{k,1} & d_{k,2} & d_{k,3} & \dots & 0 \end{bmatrix}$$

where $d_{i,j}$ represents the Jensen-Shannon distance between topic $i$ and topic $j$.

We can represent this graphically (see Figure 1) by performing a principle component analysis to scale the intertopic distances into a two-dimensional space and plotting the topics within this two-dimensional representation [Sievert et al. 2014]. From a purely qualitative view, we find that during times of recessions, topics tend to be closely clustered together, while during times of economic expansion, topics will be sparser. This is in line with the notion that no news is good news–the media is less likely to report on positive occurrences and periods of economic acceleration. However, in the face of a recession, there is a high likelihood that much media attention will be devoted to covering the economic downturn.

From an intuitive standpoint, the Jensen-Shannon distances represent how similar the topic distributions in a given time period are to one another. By taking the standard deviation of the entries within the matrix:

$$\sigma_t^{dist} = \sqrt{\frac{1}{K^2} \sum_{j=1}^{K} \sum_{i=1}^{K} (d_{i,j} - \bar{d})^2}, \qquad (4)$$

we are given a proxy for the sparsity of the given topics in a time splice. We will refer to this as the level of "coherence" within the news in a time period–the more closely related the topics are, the higher the coherence, and the more sparsely related the topics are, the lower the coherence.

We weight our coherence measure with the positive/negative scores to obtain the following metric:

$$sent_t = \sigma_t^{dist} * score_t \qquad (5)$$

This weighted metric is plotted in Figure 2. A pattern that is discernible is that in periods prior to a recession, the metric hits a local maxima, before quickly declining. This occurs prior to the beginning of the recession. From the estimated densities of the indicator, we can observe that conditional on whether or not we are in a recession period, the indicator tends to have a lower mean during times of recessions than during times of non-recession (see Figure 4). A two sample t-test and a Kolmogorov-Smirnov test both returned consistent results that the two sample are statistically different, both in means and in distribution.

## 2.3 Relative vs. Absolute Measures of Sentiment

Something that our metric does not account for is the inherent changes in the overall trends of news sentiment through time. Past studies have shown that in general, the New York Times was fairly negative from the 1960's through the early 1970's. The newspaper then became more positive in nature through the 90's, before becoming increasing negative through contemporary periods (see Leetaru, 2011; Pinker, 2018). These shifts in news sentiment are typically attributed to gradual changes in broad cultural trends through time.

We hypothesize that it is not necessarily the absolute measure of sentiment that is important, but rather, the *relative* measure of news sentiment. We construct a second measure of news sentiment, derived from our original measure of sentiment. However, we use a two year rolling window to calculate the z-score of sentiment at a given point in time:

$$z_t^{sent} = \frac{sent_t - \mu_{t-24,t}}{sd_{t-24,t}}, \tag{6}$$

where $\mu_{t-24,t}$ and $sd_{t-24,t}$ is the average sentiment score and standard deviation of the scores respectively from time $t-24$ to $t$ (as we are dealing with monthly periods, this is equivalent to 2 years). This provides us with a relative measure for sentiment in the context of the previous two year period. This allows us to normalize for larger cultural trends which may induce more negative news or more positive news.

## 3 Model Specification

We define a recession period consistent with the National Bureau of Economic Research (NBER)'s classification of a recession.[1] More specifically, we construct a binary indicator series where:

$$rec_t = \begin{cases} 1 & \text{if recession period} \\ 0 & \text{else.} \end{cases} \tag{7}$$

For simplicity, we use a standard probit model, where the expected probability of being in a recession period in a future time period is formally defined as:

$$P(rec_t = 1 | H_{t-h}) = \Phi(\pi_t), \tag{8}$$

where $H_{t-h}$ is all relevant historical information through time $t - h$, $\Phi(\cdot)$ is the standard normal cumulative distribution function, and $\pi_t$ is a function of variables.

To control for the predictive capabilities of macroeconomic and financial variables, we use the idea of common factors from Stock and Watson, 2002. To briefly summarize, the method assumes that given a panel of $T \times N$ variables made up of $N$ total individual covariance stationary series $x_{it}$ (where $i = 1, ..., N$, $t = 1, ..., T$), $x_{it}$ will follow a factor model:

$$x_{it} = \Lambda_i' F_t + \varepsilon_{it}, \tag{9}$$

---

[1] This follows from Berge and Jord, 2011 and Liu and Moench, 2016, who argued that while many other studies have been dedicated to finding leading indices to classify recessions, there existed little support for statistically significant improvement over the NBER dates using these alternative methods. Therefore, we opt to use the NBER classification.

where $F_t$ is a $k \times 1$ vector containing the common factors, $\Lambda_i$ is a $k \times 1$ vector containing factor loadings for the $i$-th variable, and $\varepsilon$ represents the idiosyncratic error. Assuming $k$ is smaller than $N$, we can effectively reduce the dimension of extremely large panel datasets.

In particular, we use a balanced panel of 138 series compiled by the Federal Reserve (see McCracken and Ng, 2015 for details) from 1965 to 2017, the makeup of which has been shown to be similar to the original Stock and Watson dataset. Linear interpolation was applied to account for missing values. We then apply principal component analysis across the panel dataset and retain 15 of the principal components to serve as our common factors.[2] We refer to the 15 common factors as $f_{i,t}$, where $i = 1, ..., 15$ and $t$ represents the time period.

To control for already used sentiment measures, we also include a set of sentiment variables proposed previously, consisting of Michigan's Index of Consumer Sentiment ($mics_t$) and the Purchasing Manager's Index ($pmi_t$).

Our proposed model takes on the following form:

$$P(rec_t = 1) = \alpha_t + \sum_{i=1}^{15} \beta_i f_{i,t-h} + \phi_1 mics_{t-h} + \phi_2 pmi_{t-h} + \gamma_1 z_{t-h}^{sent}(1 + \gamma_2 mics_{t-h} + \gamma_3 pmi_{t-h}). \tag{10}$$

We hypothesize that there exists some sort of feedback mechanism between individual sentiment (captured by $mics_t$ and $pmi_t$) and the overall news sentiment in a given time period. In particular, as more negative news is present, we expect public sentiment to decrease as well. Therefore, we introduce an interaction term between $z_t^{sent}$ and the traditional $pmi_t$ and $mics_t$.

We are interested in four different forecasting horizons: one month ahead ($h = 1$), three months ahead ($h = 3$), six months ahead ($h = 6$), and one year ahead ($h = 12$). From the regression output, we can observe that at longer forecast horizons, news sentiment and the interaction terms are highly statistically significant. However, as the forecast horizon becomes shorter, the statistical significance decreases. Furthermore, we observe that the interaction term between news sentiment and the Michigan Index of Consumer Sentiment shows that as news becomes more negative, given that consumer sentiment is also negative, the effect of negative consumer sentiment is larger than previously estimated (see Table 1

---

[2]The number of principal components to keep was determined using Bai and Ng, 2002's findings.

for full regression output).

Our primary two benchmarks are (1) a recession model that uses only the fifteen common factors:

$$P(rec_t = 1) = \alpha_t + \sum_{i=1}^{15} \beta_i f_{i,t-h}, \tag{11}$$

and (2) the common factor model, but with the inclusion of $mics_t$ and $pmi_t$:

$$P(rec_t = 1) = \alpha_t + \sum_{i=1}^{15} \beta_i f_{i,t-h} + \phi_1 mics_{t-h} + \phi_2 pmi_{t-h}. \tag{12}$$

While we initially also compared performance of our model against the New York Federal Reserve's yield curve recession model, the fifteen common factor consistently outperforms the spread (despite introducing more complexity). Therefore, we opt to simply benchmark against the fifteen common factor model as our most restricted model.

Because the two benchmarks are essentially nested models within our proposed model, we begin by conducting a series of F-tests to see whether or not adding in more factors to the base model of 15 common factors provides a statistically significant improvement in model fit for the tradeoff of adding in higher degrees of complexity. We observe that in the case of a year ahead forecast horizon, the inclusion of the Michigan Index of Consumer Sentiment and Purchasing Manager's Index does not provide a statistically significant improvement over the baseline 15 common factor model. However, our proposed model does. Therefore, we hypothesize that at a forecast horizon of one year, while consumer and producer sentiment fails to provide a statistically significant improvement in model fit, the inclusion of news sentiment and the proposed interaction terms does. In all cases except for a one month ahead forecasting horizon, the F-test returned statistically significant results in our proposed model's fit over the traditional survey-based sentiment model's fit.

A comparison of the Akaike information criterion (AIC) and Bayesian information criterion (BIC) between the models show that at longer forecast horizons (e.g., $h = 6$, $h = 12$), our proposed model is preferred; however, at shorter forecast horizons (e.g., $h = 1$, $h = 3$), AIC and BIC converge to select the 15 factor model with traditional sentiment indicators (see Table 2 for AIC and BIC values). This is in line with what we observed in our regression output, in which we saw that as the forecast horizon became shorter, the statistical significance of the relative news sentiment metric and the interaction terms decreased.

# 4 Forecasting Performance

We assess the model performance in forecasting recessions for both in-sample and out-of-sample. To measure accuracy, we apply a threshold of $C^*$ to the estimated probabilities, such that for any estimated probability greater than or equal to $C^*$, we assign a value of 1 to that time period. For all other estimated probabilities less than $C^*$, we assign a fitted value of 0. In this manner, we can measure the number of correctly classified periods.

For simplicity, the threshold of $C^*$ is set at 0.5. While we could vary this to see how the results change, the majority of our fitted and predicted probabilities by the various models fall either far below or above the threshold. We provide more discussion on the variation of this threshold in our later analysis with AUROC curves in the following section.

Because the data are inherently unbalanced–there exist more non-recession periods than recession periods–we cannot use the pure accuracy rates of each model. (More specifically, in the case of our full data set, we have 83 total recession periods, and 531 non-recession periods. This means that if we were to simply construct a model that predicted every period to be a non-recession period, we would obtain a pragmatically useless model with pure accuracy rate of around 86%.) Therefore, we use a different scoring metric known as the $F_1$-score, which is a combination of the precision and recall:

$$F_1 = 2 * \frac{precision * recall}{precision + recall}. \tag{13}$$

## 4.1 In-Sample Performance

For in-sample performance, we look at the fitted recession probabilities of each model at the different forecasting horizons. We find that for all horizons, our proposed model has the highest $F_1$-score. However, the difference in accuracy between our model and the common factor model with traditional sentiment indicators is marginal for lower forecasting horizons. The accuracy differential becomes more apparent at the longer forecast horizons (see Table 3). Visually, we can observe that the in-sample performance of our model at a forecast horizon of six months provides more stable estimates of recession probability in the last two decades in comparison to the other two models (see Figure 5 for plots).

## 4.2 Out-of-Sample Performance

To test the out-of-sample performance of the models, we partition the data into training and testing sets. By standard practice, we take the first two-thirds of the data set as the training set and use the last third of the data set as the testing set. In doing so, we find that for all forecast horizons except for $h = 1$, our proposed model performs best. We also conduct Diebold-Mariano tests between the different forecasts and find that with the exception of the one month ahead forecast, our proposed model yields statistically significantly lower error (at $\alpha = 0.05$) in the out-of-sample performance.

To test for additional model robustness, we also examine the model's capability to perform in earlier periods of time as well. To do so, we also run a sequence of backtesting by partitioning the data set such that the first third of the data are used as the testing set, and train the model on the last two-thirds of data. Additionally, we also run a sequence of backtesting where we use the first third and last third of the data as the training set, and use the middle third as the testing set. This roughly partitions our data set into three backtesting periods: 1965-1981, 1982-1998, and 1999-2017. This allows us to examine the factor's capability to perform through time in vastly different economic environments through the entire 1965 to 2017 period.

We find that within the first backtesting period, the common factor model tends to perform best. The inclusion of the sentiment factors (for our proposed metric and the survey-based metrics) seem to either provide little outperformance over using the simpler, common factor model, or provides too much additional noise, leading to a lower accuracy rate in the out-of-sample performance of the other models. Therefore, it seems that the benefit of including sentiment measures is mostly seen after the 1980's, and the benefit of including our measure of news sentiment is mostly observed after the beginning of the 21st century.

For a graphical comparison for model performance, we use the receiver operating characteristic curve (ROC curve). This is done by plotting the true positive rate against the false positive rate, with a varied probability threshold $C^*$, where $C^*$ is within the bounds of $(0, 1)$. The true positive rate is computed from how many correct positive results are produced relative to the total number of positive samples are present in the testing set.

The false positive rate is computed from how many incorrectly produced positive results are forecasted, relative to all the negative samples in the testing set. Ideally, the ROC curve of a model that forecasts 100% correct results would be equivalent to a singular point at the coordinate (0,1) (see Jorda and Taylor, 2011, 2012, Khandani et al., 2010, Liu and Moench, 2016 for further discussions of ROC curves within the context of recession predictions).

The area under the ROC curve (known as AUROC) is also a metric for model performance. To measure the amount of uncertainty associated with the estimates to determine how robustly our model performs against the others, we perform block bootstrapping across our data 200 times and recursively backtest the performance of the models through time to obtain an AUROC value. We find that consistent with what we observed from the $F_1$-score metric, our proposed model provides the most improvement in performance at longer time horizons (see Figure 7 for a visualization and Table 4 for values).

## 5 Conclusion and Future Work

To summarize, constructing a relative sentiment indicator from using estimated Jensen-Shannon distances from the Latent Dirichlet Allocation algorithm and sentiment scores allows us to construct a real-time estimate for sentiment within news data. The inclusion of such a metric allows for a significant improvement in the performance of recession prediction models containing traditional macroeconomic and financial indicators. We test our model across different forecasting horizons and different time periods and observe that the performance of recession prediction models in contemporary periods can be improved with the inclusion of a news sentiment indicator.

There exists a great deal of value that can be added to recession predictions if we are able to capture public perception and sentiment in different ways. Some future work in this area could include conducting a broader analysis using combined article text from other media sources. More specifically, our study currently only uses articles from the New York Times. It would be interesting to include other news outlets (i.e., as CNN, Fox News, MSNBC, etc.) to reduce any potential reporting bias within the New York Times. Additionally, the focus in this particular study is on the United States economy. Our methods could easily be generalized to study other country's economies, such as Germany, the United Kingdom, and Australia. LDA can operate across non-English based documents, and can even be extended to parse across multilingual documents (see Mimno, et. al 2009). Additionally, the Michigan Index of Consumer Sentiment and Purchasing Manager's Index are both proxies for public sentiment (both from a consumer and a production standpoint). However, these are not direct measures for sentiment, as they rely on surveying the public. Using similar techniques, we could construct direct measures of sentiment by utilizing social media data, such as Twitter.

# 6 Figures and Tables

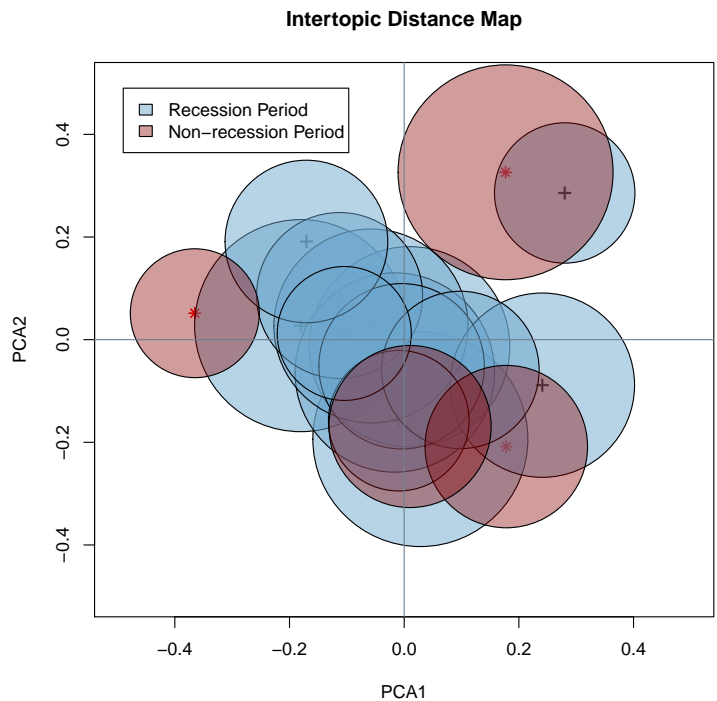

Figure 1: We compare the intertopic distances in their two-dimensional representations found for topics in a recession period and topics in a non-recession period. For the purposes of this plot, we used intertopic distances estimated for the news articles from the month of February 2009 as an example of intertopic distances within a recession period, and we used the distances estimated from the month of August 2004 as an example of intertopic distances within a non-recession period. We can observe in this graphical representation that the topics estimated during the non-recession period are much sparser than the topics estimated for the recession period.

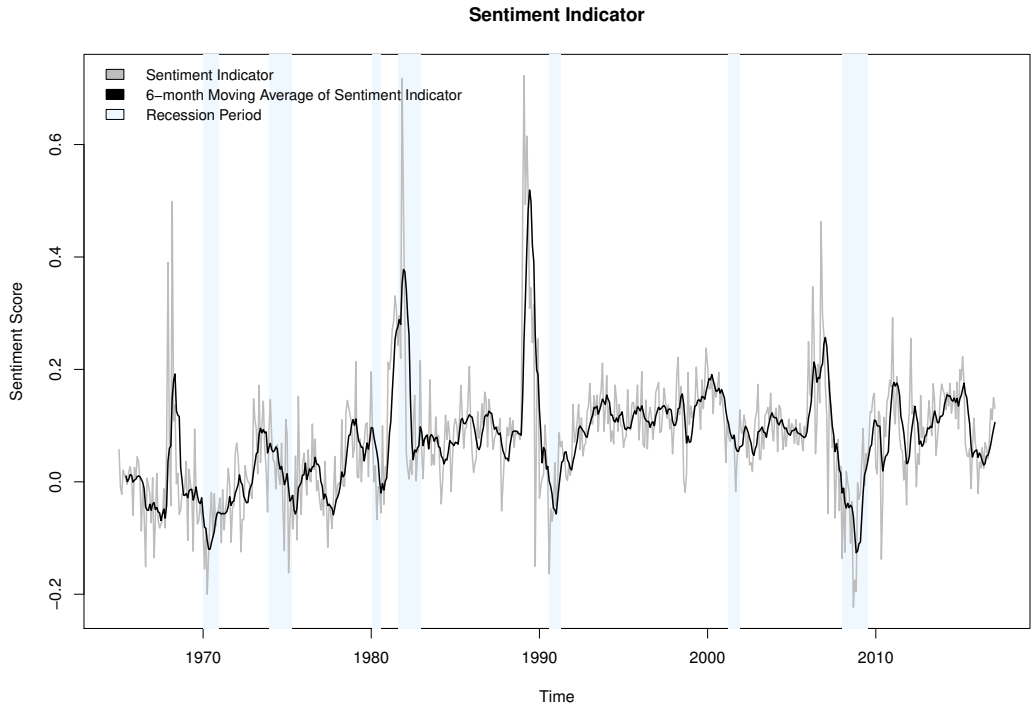

Figure 2: We plot out the constructed time series of news sentiment and observe that prior to recessions occurring (denoted by the light blue bars in the plot), the series peaks and then quickly begins declining, serving as a signal to a potential economic downturn in the future.

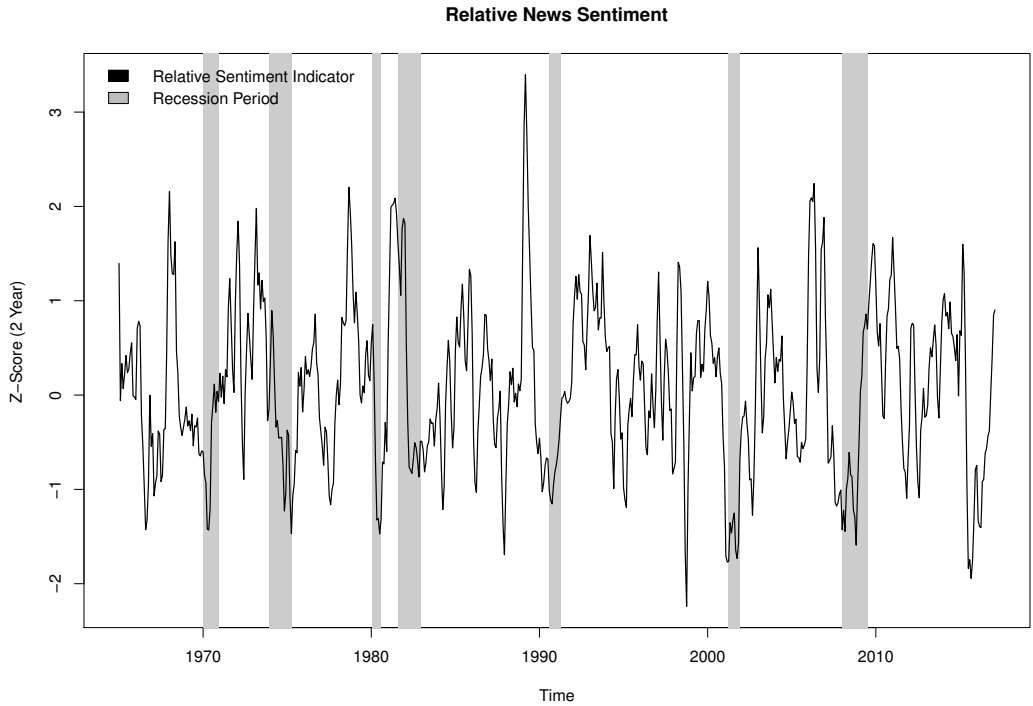

Figure 3: We plot out the relative news sentiment index that we have constructed. This is done by taking a rolling two year z-score across our original measure of news sentiment (see Figure 2 for comparison), effectively normalizing the measure of sentiment to its prior two year period. This allows us to bypass fluctuations in news cycle trends that may lead to a negative or positive bias in different periods of time.

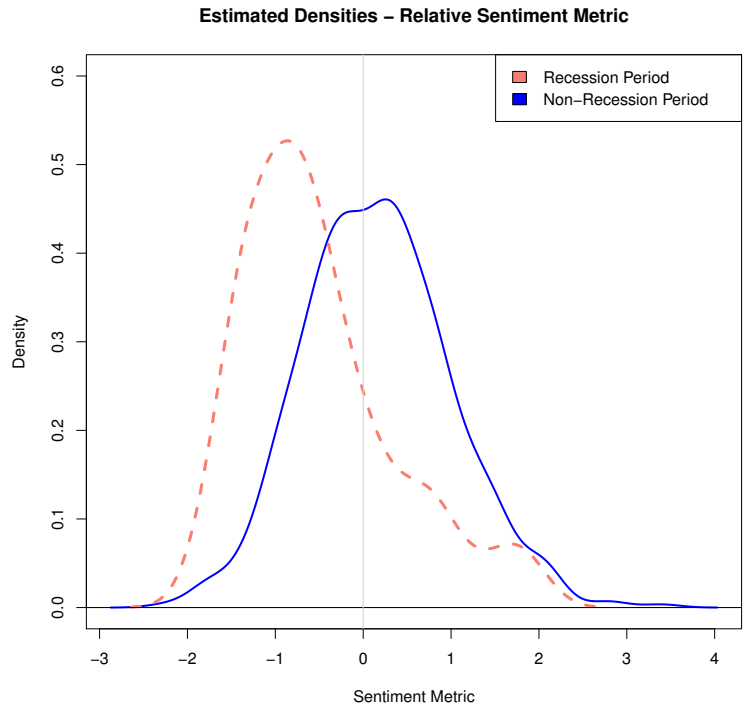

Figure 4: The estimated densities of the sentiment metric during times of recession and times of non-recession are plotted (where the dashed line notes the density during times of recession, and the solid line represents the density during times of non-recessions). We can observe that consistent with intuition, qualitatively that during recession periods, the sentiment is lower overall within the media, while during non-recession periods, the sentiment is higher.

## In-Sample Estimated Recession Probabilities

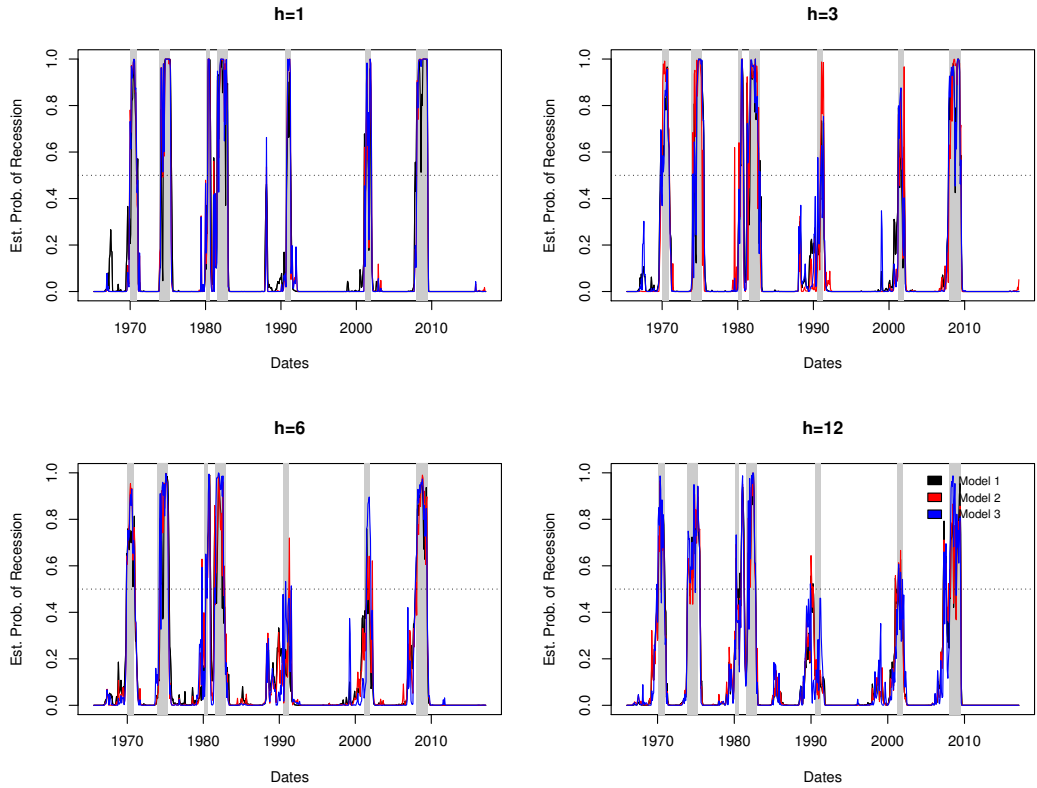

Figure 5: We compare the fitted recession probabilities for the different models. Black represents the values fit by the fifteen common factor model, red represents the values fit by the fifteen common factor and traditional sentiment indicator model, and blue represents our proposed model. We can see that the inclusion of the news sentiment indicator and the interaction terms provides a better fit. This is quantified in Table 3.

## Out-of-Sample Estimated Recession Probabilities

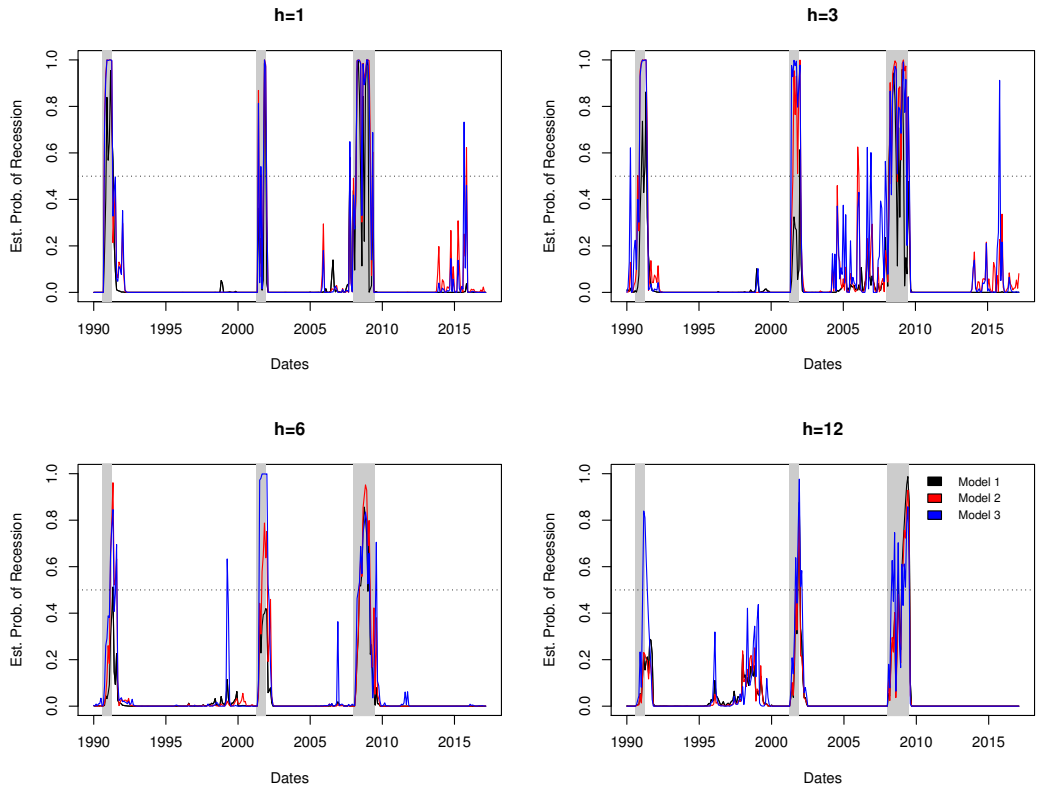

Figure 6: We plot the recursively forecasted values. We see that at a shorter time horizon, our model is prone to false signals, but is more stable and less noisy at longer time horizons. The accuracy rates of all of these models at the varying time horizons are provided in Table 3.

## Bootstrapped ROC Curve Comparison

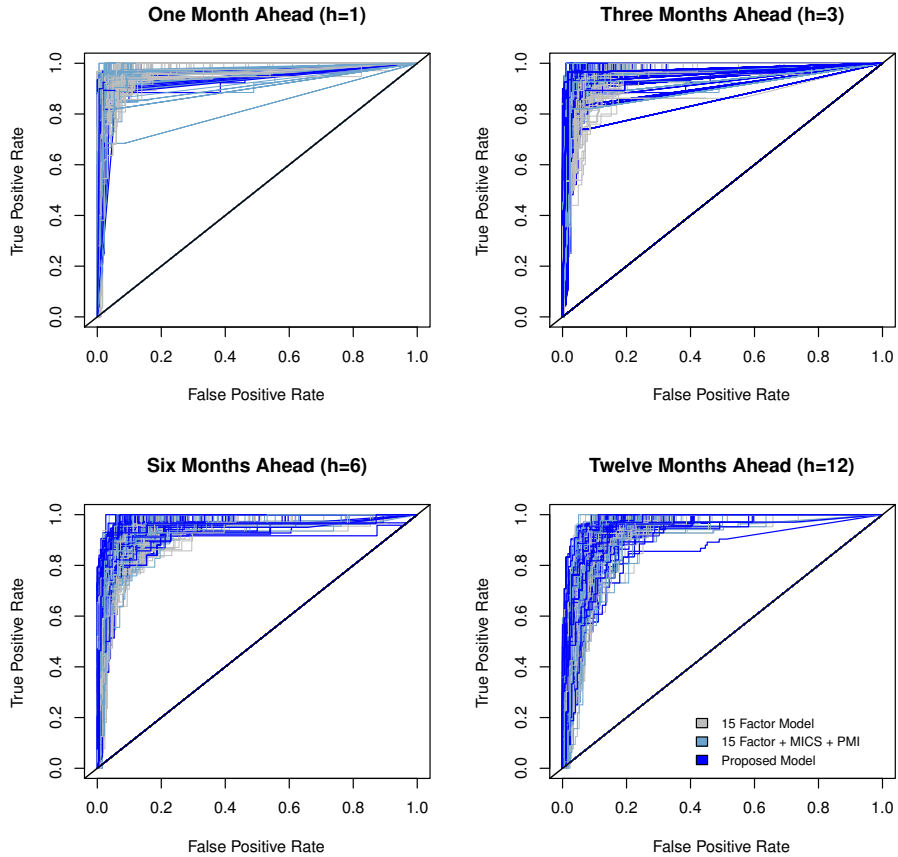

Figure 7: We perform block bootstrapping across our data set and then perform out-of-sample recursive backtesting across the last third of the synthetic data set to obtain different ROC curves. This allows us to obtain a more robust measure for how well the models perform. We observe that consistent with what we found before, the proposed model's performance is better for the longer time horizons, specifically at $h = 6$. There exists more fluctuation in its performance at $h = 3$ and $h = 12$, while at $h = 1$, it consistenly underperforms the other two models. See Table 4 for values of AUROC.

**Regression Results**

|  | *Dependent variable:* | | | |
|---|---|---|---|---|
|  | $rec_t$ | | | |
|  | $h = 1$ | $h = 3$ | $h = 6$ | $h = 12$ |
|---|---|---|---|---|
| $mics_{t-h}$ | −2.469*** | −2.265*** | −0.857*** | 0.719** |
|  | (0.540) | (0.509) | (0.287) | (0.308) |
| $pmi_{t-h}$ | 1.317*** | 1.301*** | 0.478* | 0.171 |
|  | (0.454) | (0.380) | (0.266) | (0.278) |
| $z^{sent}_{t-h}$ | −0.451 | −0.938** | −0.047 | −0.443*** |
|  | (0.499) | (0.450) | (0.184) | (0.149) |
| $z^{sent}_{t-h} * mics_{t-h}$ | −0.531 | −1.083*** | −0.842*** | −0.470*** |
|  | (0.381) | (0.370) | (0.172) | (0.135) |
| $z^{sent}_{t-h} * pmi_{t-h}$ | −0.160 | 0.511** | 0.398*** | 0.189 |
|  | (0.287) | (0.222) | (0.143) | (0.126) |
| $f_{1,t-h}$ | −0.085* | −0.064 | −0.047 | −0.769*** |
|  | (0.046) | (0.040) | (0.041) | (0.182) |
| $f_{2,t-h}$ | 0.452*** | 0.405*** | 0.243*** | 1.203*** |
|  | (0.113) | (0.093) | (0.078) | (0.266) |
| $f_{3,t-h}$ | −0.052 | −0.068 | −0.083 | −0.867*** |
|  | (0.102) | (0.106) | (0.077) | (0.244) |
| $f_{4,t-h}$ | 0.410*** | 0.490*** | 0.385*** | −0.210 |
|  | (0.111) | (0.117) | (0.105) | (0.173) |
| $f_{5,t-h}$ | 0.649*** | 0.588*** | 0.299** | 0.836*** |
|  | (0.181) | (0.171) | (0.133) | (0.238) |
| $f_{6,t-h}$ | 0.682*** | 0.438** | 0.296 | 2.066*** |
|  | (0.225) | (0.207) | (0.216) | (0.489) |
| $f_{7,t-h}$ | −2.604*** | −1.764*** | −0.742*** | 1.492*** |
|  | (0.538) | (0.422) | (0.266) | (0.446) |
| $f_{8,t-h}$ | 0.561*** | 0.537*** | 0.507*** | 2.273*** |
|  | (0.203) | (0.181) | (0.148) | (0.508) |
| $f_{9,t-h}$ | −0.057 | −0.345 | 0.062 | 0.640** |
|  | (0.281) | (0.255) | (0.166) | (0.253) |
| $f_{10,t-h}$ | 0.278 | −0.366 | −0.129 | −1.102*** |
|  | (0.378) | (0.310) | (0.225) | (0.276) |
| $f_{11,t-h}$ | 1.678*** | 1.314*** | 0.393 | 0.397 |
|  | (0.508) | (0.441) | (0.248) | (0.336) |
| $f_{12,t-h}$ | −1.715** | −0.101 | −0.099 | 0.433 |
|  | (0.672) | (0.552) | (0.430) | (0.581) |
| $f_{13,t-h}$ | 0.176 | 0.367 | 0.707*** | 2.482*** |
|  | (0.330) | (0.258) | (0.209) | (0.432) |
| $f_{14,t-h}$ | −0.209 | 0.464 | 0.510* | −0.275 |
|  | (0.547) | (0.428) | (0.305) | (0.338) |
| $f_{15,t-h}$ | −0.855 | −1.391** | −1.047*** | −0.381 |
|  | (0.539) | (0.550) | (0.321) | (0.318) |
| Constant | −5.630*** | −5.217*** | −2.941*** | −6.893*** |
|  | (0.904) | (0.855) | (0.362) | (1.255) |
| Observations | 625 | 623 | 620 | 614 |
| Log Likelihood | −39.365 | −49.709 | −77.942 | −98.311 |
| Akaike Inf. Crit. | 120.730 | 141.419 | 197.884 | 238.621 |

Model Selection Criteria

|  | df | h = 1 | | h = 3 | | h = 6 | | h = 12 | |
| --- | --- | --- | --- | --- | --- | --- | --- | --- | --- |
| Model | | AIC | BIC | AIC | BIC | AIC | BIC | AIC | BIC |
| Model 1 (15 PCA) | 16 | 152.2288 | 223.2328 | 194.1055 | 265.0582 | 337.7354 | 308.6109 | 253.886 | 324.606* |
| Model 2 (MCI, PMI, 15 PCA) | 18 | 117.6832* | 197.5627* | 149.1068* | 228.9286* | 221.9752 | 301.7102 | 254.081 | 333.641 |
| Model 3 (Our proposed model) | 21 | 120.7297 | 213.9225 | 183.9261 | 268.1824 | 197.8835* | 290.9076* | 238.621* | 331.441 |

Table 2: We compare model specifications between a model containing only 15 common factors, a model containing the 15 common factors and the two traditional sentiment measures, and our proposed model. An asterisk denotes the best performing model at a given forecast horizon.

| **Model Performance** | | | | |
|---|---|---|---|---|
| ***In-Sample*** | h=1 | h=3 | h=6 | h=12 |
| Model 1 (15 PCA) | 0.8571 | 0.7875 | 0.7200 | 0.6710 |
| Model 2 (MCI, PMI, 15 PCA) | 0.9091 | 0.8589 | 0.7564 | 0.6795 |
| Model 3 (Our proposed model) | 0.9146 | 0.8606 | 0.7975 | 0.7355 |
| | | | | |
| ***Out-of-Sample*** | | | | |
| **Period 1** | h=1 | h=3 | h=6 | h=12 |
| Model 1 (15 PCA) | 0.6813 | 0.7865 | 0.8409 | 0.4390 |
| Model 2 (MCI, PMI, 15 PCA) | 0.6392 | 0.7957 | 0.8506 | 0.4444 |
| Model 3 (Our proposed model) | 0.5769 | 0.7912 | 0.7955 | 0.5063 |
| **Period 2** | h=1 | h=3 | h=6 | h=12 |
| Model 1 (15 PCA) | 0.7143 | 0.4545 | 0.0000 | 0.0000 |
| Model 2 (MCI, PMI, 15 PCA) | 0.8966 | 0.6154 | 0.2222 | 0.0000 |
| Model 3 (Our proposed model) | 0.8000 | 0.5000 | 0.2500 | 0.1250 |
| **Period 3** | h=1 | h=3 | h=6 | h=12 |
| Model 1 (15 PCA) | 0.5000 | 0.5789 | 0.4706 | 0.4444 |
| Model 2 (MCI, PMI, 15 PCA) | 0.7273 | 0.7660 | 0.6500 | 0.4444 |
| Model 3 (Our proposed model) | 0.6977 | 0.8077 | 0.6818 | 0.5000 |

Table 3: $F_1$-scores for both in-sample and out-of-sample performance. We observe that at longer time horizons, our proposed model has greater outperformance in accuracy, relative to the other models. We also find that much of this outperformance occurs during the last period (i.e., the last two decades).

**AUROC Comparison**

|  |  | Min | 1st Quartile | Median | Mean | 3rd Quartile | Max |
|---|---|---|---|---|---|---|---|
| $h = 1$ | (1) | 0.939 | 0.975 | 0.982 | 0.980 | 0.988 | 0.999 |
|  | (2) | 0.818 | 0.964 | 0.982 | 0.974 | 0.992 | 0.999 |
|  | (3) | 0.878 | 0.955 | 0.968 | 0.966 | 0.988 | 1.000 |
| $h = 3$ | (1) | 0.872 | 0.964 | 0.973 | 0.970 | 0.979 | 0.990 |
|  | (2) | 0.890 | 0.977 | 0.985 | 0.981 | 0.989 | 0.995 |
|  | (3) | 0.854 | 0.976 | 0.984 | 0.977 | 0.989 | 0.999 |
| $h = 6$ | (1) | 0.893 | 0.943 | 0.956 | 0.955 | 0.968 | 0.987 |
|  | (2) | 0.913 | 0.952 | 0.962 | 0.961 | 0.972 | 0.986 |
|  | (3) | 0.906 | 0.959 | 0.972 | 0.968 | 0.979 | 0.987 |
| $h = 12$ | (1) | 0.887 | 0.938 | 0.949 | 0.948 | 0.958 | 0.986 |
|  | (2) | 0.889 | 0.939 | 0.950 | 0.949 | 0.961 | 0.986 |
|  | (3) | 0.901 | 0.947 | 0.958 | 0.956 | 0.966 | 0.988 |

Table 4: Distribution of AUROC values from bootstrapping the data. We denote (1) as the model containing only 15 common factors, (2) as the model with 15 common factors, $mics_t$, and $pmi_t$, and (3) as our proposed model with all 15 common factors, $mic_t$, $pmi_t$, and our metric, along with interaction terms. We see that at longer time horizons (i.e., $h = 6$, $h = 12$), the average AUROC performance is greater for our proposed model.

[24] Diebold, F. & Li, C., "Forecasting the term structure of government bond yields", *Journal of Economics*, 130, 337-364, (2006)

[25] Diebold, F., Rudebusch, G. D., & Aruoba, S. B., "The macroeconomy and the yield curve: a dynamic latent factor approach", *Journal of Econometrics*, 131, 309-338, (2006)

[26] Estrella, A. & Mishkin, F. S., "Predicting U.S. Recessions: Financial Variables as Leading Indicators", *Review of Economics and Statistics, Vol. 80, No. 1: pp. 45-61*, (1998)

[27] Estrella, A., Rodriguez, A. P., & Schich, S., "How stable is the predictive power of the yield curve: evidence from Germany and the United States", *Review of Economics and Statistics*, 85, (2003)

[28] Estrella, A. & Trubin, M. R., "The Yield Curve as a Leading Indicator: Some Practical Issues", *Current Issues in Economics and Finance, Vol. 12, No. 5*, (2006)

[29] Gilbert, E. & Karahalios, K., "Widespread Worry and the Stock Market", *Proceedings of the Fourth International AAI Conference on Weblogs and Social Media,* (2009)

[30] Gogas, P., Papadimitriou, T., Matthaiou, M., & Chrsyanthidou, E., "Yield Curve and Recession Forecasting in a Machine Learning Framework", *Computational Economics*, Volume 45, Issue 4, pp 635-645, (2015)

[31] Hisano, R., Sornette, D., Mizuno, T., & Ohnishi, T., "High quality topic extraction from business news explains abnormal financial market volatility", *PLoS ONE 8(6): e64846*, (2013)

[32] Jordà, Ó., & Taylor, A. M., "Performance evaluation of zero net- investment strategies", Working Paper 17150, *National Bureau of Economic Research*, (2011)

[33] Jordà, Ó., & Taylor, A. M., "The carry trade and fundamentals: Nothing to fear but fear itself", *Journal of International Economics*, 88(1), 74-90, (2012)

[34] Koenig, E., "Using the purchasing managers' index to assess the economy's strength and the likely direction of monetary policy", *Economic and Financial Policy Review*, Federal Reserve Bank of Dallas, 1 (6), 1-14, (2002)